# Turbulence is ineffective in causing raindrop growth in polluted clouds

K. Shri Vignesh[1,2], Ambedkar Sanket Sukdeo[1,2], P. V. Sruthibhai[1,2], Aishwarya Singh[3,4,5], Srikrishna Sahu[6], Swetaprovo Chaudhari[7], Amit K. Patra[8], T. Narayana Rao[8], Rama Govindarajan[9*], Sachin S. Gunthe[3,4], and R. I. Sujith[1,2*]

[1]Department of Aerospace Engineering, Indian Institute of Technology Madras, Chennai, India

[2]Centre of Excellence for Studying Critical Transitions in Complex Systems, Indian Institute of Technology Madras, Chennai, India

[3]Environmental Engineering Division, Department of Civil Engineering, Indian Institute of Technology Madras, Chennai, India

[4]Centre for Atmospheric and Climate Sciences, Indian Institute of Technology Madras, Chennai, India

[5]Department of Aerosol Chemistry, Max Planck Institute for Chemistry, Mainz, Germany

[6]Department of Mechanical Engineering, Indian Institute of Technology Madras, Chennai, India

[7]Institute for Aerospace Studies, University of Toronto, Toronto, Canada

[8]National Atmospheric Research Laboratory, Gadanki, India

[9]International Centre for Theoretical Sciences, Tata Institute of Fundamental Research, Bangalore, India

To whom correspondence should be addressed: R. I. Sujith (sujith@iitm.ac.in) & Rama Govindarajan (rama@icts.res.in)

**Abstract:**

Aerosol-cloud interactions represent the largest uncertainty in climate-change assessment, and while cloud turbulence is considered crucial for droplet growth, its precise role remains unclear. Our laboratory-controlled studies show that turbulence does not always enhance collision and coalescence; instead, its influence emerges only when droplets have a sufficiently broad size distribution. The dissipative-scale droplet behaviour underscores the importance of improved parameterisations to accurately model cloud microphysics.

Clouds, play a crucial role in regulating radiative forcing and precipitation, thereby exerting a significant influence on the hydrological cycle[1]. Variations in the characteristics and properties of clouds in response to increased anthropogenic aerosol levels represent a prominent source of uncertainty in forecasts related to climate impacts modulated by clouds[2]. This uncertainty is particularly pronounced in predicting the onset of precipitation, primarily due to an insufficient understanding of the underlying microphysical processes involved[3]. While our comprehension of the processes linking changes in aerosol properties to the formation of cloud droplets have advanced over the past decade[4], substantial challenges remain in elucidating the mechanisms that govern the rapid transition of cloud droplets to rain droplets[3,5]. The transition from condensation-dominated to turbulence-dominated droplet growth is critical, as turbulence facilitates collision and coalescence processes that ultimately lead to gravity-driven runaway growth[5-7].

Due to the limited understanding of the role of turbulence in droplet growth, current climate and weather models employ microphysical parameterizations that rely heavily on process models, which tend to oversimplify the essential aspects of cloud microphysics and do not adequately account for the intricate interplay between the initial droplet size distribution and the impact of turbulence on collision and coalescence processes[8-9]. Cloud droplet size distributions are strongly influenced by aerosol concentration, with polluted conditions producing narrower distributions and higher droplet number concentrations than those formed under clean or pristine conditions[10-13]. Crucially, the degree to which turbulence modulates droplet growth in varying aerosol environments and different local droplet size distributions within a cloud[14] remains poorly characterised.

Laboratory-controlled experiments offer the advantage of maintaining consistent conditions, enabling accurate measurements of the relevant parameters while preserving the physical principles at play[15], thus providing valuable insights into the initiation of collision and coalescence under varying turbulent conditions and aerosol environments. A deeper understanding of the microphysical processes can improve our understanding and prediction of the onset of precipitation and the accurate assessment of the lifetime of a cloud, thereby improving our assessment of the radiative forcing due to cloud albedo.

Here, through a specialised laboratory experiment, we investigated the influence of turbulence on three droplet size distributions characteristic of those observed in clouds under varying initial aerosol concentrations from polluted to pristine environments. Clouds originating in polluted conditions are characterized by a narrow droplet size distribution and higher number concentration, while clouds formed in relatively clean conditions and pristine environments are characterized by a broader droplet size distribution and fewer droplets (Extended Data Fig. 3). We demonstrate conclusively over a range of conditions that turbulence is ineffective in causing any noticeable droplet growth when their size distribution is narrow. A broad size distribution is essential, which includes larger droplets which are prone to rapid growth, and smaller droplets which the larger ones may prey on.

Droplets with appropriate distributions were introduced into the chamber at a constant rate where homogeneous isotropic turbulence was generated. Turbulence levels are commonly reported using Reynolds numbers, which represent the ratio of inertial to viscous forces. Since small-scale flow structures are most relevant for droplet dynamics, we characterize the Reynolds numbers using the Taylor-microscale ($Re_\lambda = \sqrt{15 u_{rms} L / \nu}$), where $u_{rms}$ is the root-mean-square velocity of the turbulent fluctuations and $L$ is the largest length scale of the turbulence (Extended Data Fig. 2). Upon the flow reaching steady-state, the droplet size distribution and high-resolution laser-illuminated images of the droplet field were obtained (refer Methods and Extended Data). We aim to determine the conditions under which droplet growth occurs and to elucidate the underlying mechanisms driving this process.

Droplet collision and coalescence is essential to rain formation, and their actual rates are far higher than can be explained so far. The influence of turbulence on droplet collision enhancement strongly depends on the droplet size distribution. Despite a very high number density, turbulence has no significant effect on the droplet size distribution corresponding to a polluted condition (Fig. 1a). The absence of droplet growth is striking across the range of $Re_\lambda$ over which our experiments were performed (see inset in Fig. 1a), highlighting the stability of the droplet size distributions subjected to varying turbulence levels. In contrast, we observe a shift in the droplet size distribution towards bigger droplets for distributions corresponding to relatively clean (Fig 1b) and pristine conditions (Fig 1c) with an increase in $Re_\lambda$. The shift is more pronounced for pristine conditions where the increased probability of droplets with size greater than 15 μm, accompanied by a decrease in the probability of droplets smaller than 15 μm, indicates growth driven by coalescence of smaller droplets under the influence of turbulence (see inset in Fig 1c). Note that this increase in droplet size range (15 – 40 μm) coincides with the effective sizes for enhanced drizzle formation during the onset of rain[13], underscoring the critical role of the droplet size distribution in turbulence-driven collision and coalescence processes.

To investigate the underlying mechanisms of collision enhancement by turbulence, we characterized clustering among droplets using the radial distribution function ($g(r)$), obtained from images of the droplet field (refer Methods and Extended Data). $g(r)$ quantifies droplet clustering by comparing the likelihood of finding another droplet at a separation $r$ relative to that expected for randomly distributed droplets. Figure 1d shows the variation of $g(r)$ depicted by the colour scale, with the average Stokes number ($\overline{St}$) and $r/\eta$, where $\eta$ is the Kolmogorov length scale. Here, Stokes number ($St = \tau_p/\tau_\eta$) characterizes the inertial response, indicating the ability of droplets to follow the streamlines and is estimated from the ratio of particle response time ($\tau_p = \rho D^2/18\mu$) to the Kolmogorov time ($\tau_\eta$).

For the relatively clean and the pristine conditions ($\overline{St}$ > 0.2), we observe enhanced clustering among droplets ($g(r) > 1$), with distinctly higher vales at smaller scales (see inset in Fig 1d), similar to that reported in recent studies[16-17]. Such intense clustering at the smallest scales can substantially enhance the local number densities and collision probability. Furthermore, the broad distribution characteristic of pristine conditions produces droplets with a wide range of St. (Extended Data Fig. 5). We observe that bigger droplets (St > 3) exhibit markedly lower rms velocities than the tracer-like droplets (St < 0.3), reflecting their inability to follow the rapid fluctuations of the turbulent flow (Extended Data Fig. 6). Consequently, bigger droplets retain their velocities that differ from the local fluid velocity and from neighbouring small droplets, producing large relative velocities (caustics) between droplets with disparate St[18]. The coexistence of strong clustering at the smaller scales along with relative velocities between droplets of different St implies a multiplicative enhancement in collision rates by turbulence. However, for polluted conditions ($\overline{St}$ < 0.2), $g(r)$ values for all $r/\eta$ are close to 1, indicating no appreciable clustering among droplets. Moreover, the narrow size distribution in polluted condition yields droplets with similar St, reducing the chance of caustic formation, minimising the collision enhancement.

This study elucidates a clear microphysical pathway by which turbulence can accelerate warm rain initiation, highlighting the importance of droplet size distribution obtained by condensation-driven growth. Recent laboratory and theoretical studies highlight the role of turbulence in inducing supersaturation fluctuations in producing bigger droplets under pristine conditions[19]. These studies describe how turbulence can produce the first population of larger droplets through stochastic condensation, whereas our results identify the subsequent stage in which turbulence enhances collision and coalescence. In this regime, both spatial clustering, particularly at dissipative scales, and caustic formation emerge as key mechanisms driving collision enhancement. Our experiments reveal that these mechanisms become active only when droplets span a sufficiently broad range of sizes, suggesting a possible existence of a critical distribution beyond which turbulence can influence collision and coalescence.

Since droplet size distributions are strongly influenced by aerosol concentrations at the cloud's origin, our results provide a mechanistic explanation for the observed longer lifetimes and suppressed precipitation in polluted clouds compared to their pristine counterparts (Fig. 2). The lack of clustering and turbulence insensitivity towards droplet size growth in a distribution similar to polluted clouds suggest that the narrowing of the droplet size distribution, which is a consequence of a large concentration of aerosols imposes a fundamental limitation in turbulence-driven growth. Incorporating these turbulence-induced microphysical effects into parameterizations could significantly improve the representation of warm-rain processes in climate models, reducing uncertainties in precipitation forecasting.

Furthermore, these findings bear on marine cloud brightening strategies[20], which rely on sustaining a narrow distribution of droplets to increase cloud lifetimes. Our experiments show that turbulence cannot broaden a narrow distribution despite very high droplet number densities and the stability of the distribution under varying turbulence conditions is crucial to the efficacy and potential impacts of these methods. To summarise, for turbulence to bridge the droplet growth bottleneck, a size distribution of droplets which includes both large and small droplets, as indicative of relatively clean or pristine environments, is a necessary precondition. Turbulence is likely to be ineffective in providing this bridge in polluted clouds.

**References:**

1. Baker, M. B. & Peter, T. Small-scale cloud processes and climate. *Nature* **451**, 299–300 (2008).

2. Watson-Parris, D. & Smith, C. J. Large uncertainty in future warming due to aerosol forcing. *Nat. Clim. Change* **12**, 1111–1113 (2022).

3. Stier, P. et al. Multifaceted aerosol effects on precipitation. *Nat. Geosci.* **17**, 719–732 (2024).

4. Gunthe, S. S. et al. Cloud condensation nuclei in pristine tropical rainforest air of Amazonia: size-resolved measurements and modelling of atmospheric aerosol composition and CCN activity. *Atmos. Chem. Phys.* **9**, 7551–7575 (2009).

5. Chandrakar, K. K., Morrison, H., Grabowski, W. W. & Lawson, R. P. Are turbulence effects on droplet collision–coalescence a key to understanding observed rain formation in clouds? *Proc. Natl Acad. Sci. USA* **121**, e2319664121 (2024).

6. Bodenschatz, E., Malinowski, S. P., Shaw, R. A. & Stratmann, F. Can we understand clouds without turbulence? *Science* **327**, 970–971 (2010).

7. Devenish, B. J. et al. Droplet growth in warm turbulent clouds. *Q. J. R. Meteorol. Soc.* **138**, 1401–1429 (2012).

8. Morrison, H. et al. Confronting the challenge of modelling cloud and precipitation microphysics. *J. Adv. Model. Earth Syst.* **12**, e2019MS001689 (2020).

9. Grabowski, W. W. et al. Modelling of cloud microphysics: Can we do better? *Bull. Am. Meteorol. Soc.* **100**, 655–672 (2019).

10. Rosenfeld, D. Suppression of rain and snow by urban and industrial air pollution. *Science* **287**, 1793–1796 (2000).

11. Rosenfeld, D. et al. Flood or drought: How do aerosols affect precipitation? *Science* **321**, 1309–1313 (2008).

12. Lu, M. -L. et al. The marine stratus/stratocumulus experiment (MASE): Aerosol–cloud relationships in marine stratocumulus. *J. Geophys. Res. Atmos.* **112**, D10209 (2007).

13. Glienke, S. et al. Cloud droplets to drizzle: Contribution of transition drops to microphysical and optical properties of marine stratocumulus clouds. *Geophys. Res. Lett.* **44**, 8002–8010 (2017).

14. Allwayin, N., Larsen, M. L., Glienke, S. & Shaw, R. A. Locally narrow droplet size distributions are ubiquitous in stratocumulus clouds. *Science* **384**, 528–532 (2024).

15. Shaw, R. A. et al. Cloud–aerosol–turbulence interactions: Science priorities and concepts for a large-scale laboratory facility. *Bull. Am. Meteorol. Soc.* **101**, E1026–E1035 (2020).

16. Yavuz, M. A., Kunnen, R. P. J., van Heijst, G. J. F. & Clercx, H. J. H. Extreme small-scale clustering of droplets in turbulence driven by hydrodynamic interactions. *Phys. Rev. Lett.* **120**, 244504 (2018).

17. Johnson, D. R., Hammond, A. L., Bragg, A. D. & Meng, H. Detailed characterization of extreme clustering at near-contact scales in isotropic turbulence. *J. Fluid Mech.* **982**, A21 (2024).

18. James, M. & Ray, S. S. Enhanced droplet collision rates and impact velocities in turbulent flows: The effect of polydispersity and transient phases. *Sci. Rep.* **7**, 12231 (2017).

19. Chandrakar, K. K. et al. Aerosol indirect effect from turbulence-induced broadening of cloud-droplet size distributions. *Proc. Natl Acad. Sci. USA* **113**, 14243–14248 (2016).

20. Feingold, G. et al. Physical science research needed to evaluate the viability and risks of marine cloud brightening. *Sci. Adv.* **10**, eadi8594 (2024).

**Data availability:**

All data supporting the findings of this study are provided within the article and its Extended Data. Processed data and the image-processing code used in this study are available at https://doi.org/10.6084/m9.figshare.30788024. Due to the large size of the experimental raw image files, these data are available from the corresponding author upon reasonable request.

**Acknowledgements:**

The authors gratefully acknowledge financial support from the Exploratory Research Projects, IIT Madras (No. ASE/18-19/846/RFER/RISU), the ISRO–IITM Cell (No. SP/21-22/1197/AE/ISRO/002696), and the IoE Initiative (No. SP/22-23/1222/CPETWOCTSH). We thank S. Anand, S. Thilagaraj, K. V. Reeja and G. Sudha for their assistance in setting up the experiments. We are especially grateful to the Central Electronics Centre, IIT Madras, for their support in establishing the facility. K.S.V. and S.A acknowledges the research assistantship provided by the Ministry of Human Resource Development, India, and the Indian Institute of Technology Madras.

**Author contributions:**

K.S.V., S.S., S.C., A.K.P., T.N.R., R.G., S.S.G., and R.I.S. conceived and conceptualized the study. K.S.V., S.A., S.P.V., S.C., and R.I.S. designed and developed the facility. K.S.V., S.A., and S.P.V. performed the experiments. K.S.V. wrote codes and processed the data. K.S.V., S.S., S.C., A.K.P., T.N.R., R.G., S.S.G., and R.I.S analysed the results. All authors contributed to manuscript preparation and reviewing.

**Competing interest:**

The authors declare no competing interests.

## Methods

### Experimental conditions

Experiments were conducted in a closed, fan-driven turbulence chamber[1-2], where homogeneous and isotropic turbulence is generated at the centre (Extended Data Fig. 1). Different levels of turbulence can be achieved by varying the speed of the fans (Extended Data Fig. 2). Droplets were generated using ultrasonic microporous atomizers, which consist of a stainless-steel plate perforated with micron-sized orifices driven by a piezoceramic ring[3-4]. When supplied with water and actuated at the resonant frequency, the vibrating plate produces uniform droplets whose diameters are determined primarily by the orifice size. We employed atomizers with nominal orifice diameters of 5 μm, 9 μm, and 15 μm to generate distinct droplet size distributions representative of polluted, relatively clean, and pristine conditions, respectively (Extended Data Fig. 3). Arrays of identical atomizers were operated to achieve the required droplet number densities. Input power and water flow rates were held constant across all turbulence conditions, and the atomizers were run continuously to maintain statistical stationarity.

The droplet volume fraction maintained in the turbulence chamber for all experiments was $10^{-6}$ or lower, thus the modulation of turbulence due to the droplets can be neglected. The Froude number ($\text{Fr} = \epsilon^{3/4}/(g\nu^{1/4})$), defined as the ratio between acceleration caused by turbulent flow and that of gravity[3], estimates the relative significance of gravity and turbulence. For all the experimental conditions Fr is greater than 10, indicating that gravitational effects on droplet collisions are negligible compared with turbulence[5]. This effective decoupling of gravitational influences from turbulent dynamics highlights the uniqueness of this study.

### PDPA measurements

Droplet diameters and one component of the velocity were measured using a Phase Doppler Particle Analyzer[6] (PDPA; TSI Instruments) (Extended Data Fig. 4). The system employed two intersecting laser beams at 532 nm, with the receiving optics positioned at a 33° off-axis scattering angle. The system can measure droplet sizes in the range of 0.5 - 180μm with an uncertainty of ± 0.5μm. Only droplets with sufficient signal-to-noise ratio and validated phase bursts were accepted. Diameter and velocity measurements were recorded in coincidence, and manufacturer-provided signal-validation routines (including intensity and diameter-difference checks) were applied in real time. Across all experimental conditions, the validation rate remained high (≈ 90%), ensuring robust statistics.

**Imaging and optical configuration**

High-resolution images of the droplet field were acquired with a high-speed CMOS camera operated at 1 KHz with resolution 1280 x 1024, coupled with a 100 mm macro-lens with f/2 aperture (Extended Data Fig. 4). The camera was positioned normal to the sheet, yielding a field of view of 30 mm x 23.5 mm. This spatial calibration was performed using a precision target. Laser beam from Nd:YLF laser with a wavelength 527 nm and 25 mJ per pulse was used as the light source. A laser sheet of thickness 1 mm is created using a beam collimator and a combination of a 25 mm plano-concave cylindrical lens and a 600 mm spherical plano-convex lens mounted on a light sheet optics.

**Experimental procedure**

Experiments were performed at turbulence levels greater than $Re_\lambda = 136$ to thoroughly mix the droplets throughout the chamber and to minimize the effect of the atomizers on the turbulence in the region of measurement. For every droplet size distribution, an identical set of twelve turbulence conditions was selected, with $Re_\lambda$ values ranging from 136 to 207. Each experiment followed an identical sequence. The fans were first set to the target rotation rate and allowed to stabilize. Atomizers were then switched on, and the droplets were allowed to equilibrate for 30 s. A steady-state cloud-like condition is achieved where the constant rate of droplets injected by the atomizers is balanced by the droplet growth by collision and coalescence due to turbulence, and droplets are lost by hitting the walls and by settling. Throughout the duration of the experiment, both the turbulence properties and the atomizer operational settings were kept constant. After reaching steady-state conditions, both imaging and PDPA measurements were acquired consecutively. For each turbulence level, five independent realisations were performed, and in each realization, 10000 PDPA measurements and 2000 images were acquired.

**Environmental monitoring and ancillary measurements**

Temperature, relative humidity, and laser power were continuously monitored throughout all experiments. Chamber temperature and humidity were measured using a Vaisala HMP 7 probe and remained stable throughout, with values of approximately 24 °C (± 0.1°C) and 95 % (± 1%) relative humidity. The entire setup was housed inside a temperature-humidity-controlled laboratory where the temperature and relative humidity of 24 °C (± 1°C) and 50 % (± 5%) were maintained respectively.

Fogging of windows posed a significant challenge while performing long-duration experiments. To mitigate this issue, the windows were coated with a transparent superhydrophobic material, the Glaco MirrorCoat Zero purchased from Soft99 Co.[7]. Also, externally heated hot air at 40 °C was constantly impinged on the outer surface of the windows to gently heat and maintain their surface temperature slightly above the dew point, thereby preventing condensation. This heating was very mild and did not affect the air temperatures inside the experimental setup.

**Image Processing:**

The acquired images were processed using in-house codes based on MATLAB®. The images were binarized using an optimum intensity threshold. The droplet number density was estimated by counting the number of droplets within the illumination volume. From the images, the radial distribution function $g(r)$[8] is calculated by binning the particle pairs at different separation distances $r$, based on the below formula,

$$g(r) = \frac{N_r/A_r}{N_T/A_T}$$

here, $N_r$ represents the number of droplet pairs found in the annular area $A_r$ at a distance $r$ with a separation $r \pm dr$, where $N_T$ is the total number of droplet pairs in the image with viewing area $A_T$. Periodic boundaries we assumed while calculating $g(r)$, where the particle field is mirrored across the image boundaries[1,9]. We examined the effect of thresholding on the trend of $g(r)$ by varying the threshold value by ± 20% of the selected threshold value and found the results to be qualitatively similar. The code used for image processing is available at https://doi.org/10.6084/m9.figshare.30788024.

**References:**

1. Salazar, J. P. et al. Experimental and numerical investigation of inertial particle clustering in isotropic turbulence. *J. Fluid Mech.* **600**, 245–256 (2008).

2. Dou, Z. et al. PIV measurement of high-Reynolds-number homogeneous and isotropic turbulence in an enclosed flow apparatus with fan agitation. *Meas. Sci. Technol.* **27**, 035305 (2016).

3. Chen, H. et al. Experimental study on optimal spray parameters of piezoelectric atomizer based spray cooling. *Int. J. Heat Mass Transf.* **103**, 57–65 (2016).

4. Hou, Y. et al. Recent trends in structures and applications of valveless piezoelectric pump—a review. *J. Micromech. Microeng.* **32**, 053002 (2022).

5. Bec, J., Homann, H. & Ray, S. S. Gravity-driven enhancement of heavy particle clustering in turbulent flow. *Phys. Rev. Lett.* **112**, 184501 (2014).

6. Albrecht, H. E., Damaschke, N., Borys, M. & Tropea, C. Laser Doppler and Phase Doppler Measurement Techniques. (*Springer*, Berlin, 2013).

7. Bourrianne, P., Lv, C. & Quéré, D. The cold Leidenfrost regime. *Sci. Adv.* **5**, eaaw0304 (2019).

8. Sundaram, S. & Collins, L. R. Collision statistics in an isotropic particle-laden turbulent suspension. Part 1. Direct numerical simulations. *J. Fluid Mech*. **335**, 75–109 (1997).

9. De Jong, J. et al. Measurement of inertial particle clustering and relative velocity statistics in isotropic turbulence using holographic imaging. *Int. J. Multiphase Flow* **36**, 324–332 (2010).

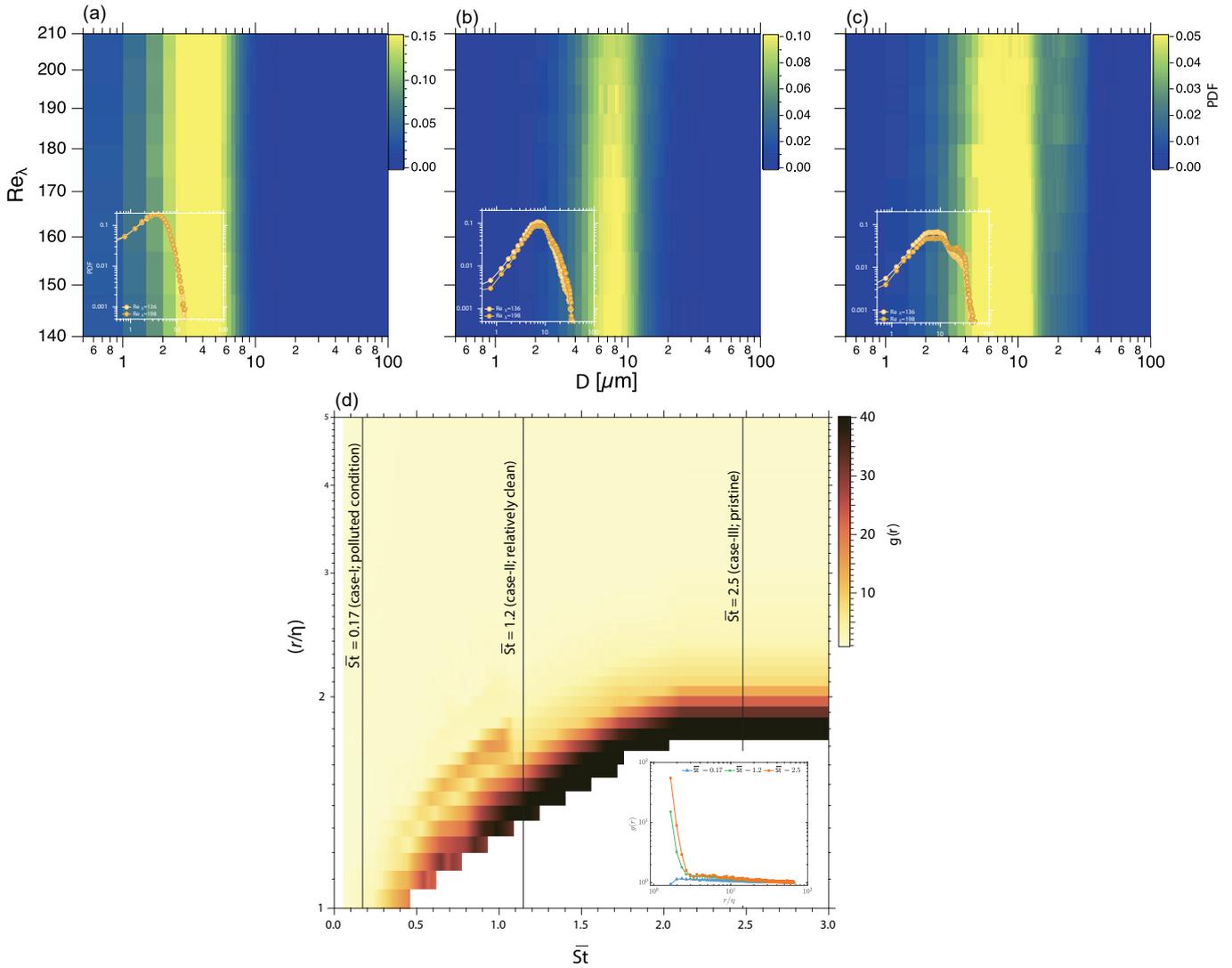

**Figure 1: Effect of turbulence on cloud droplet size growth: (a), (b),** and **(c)** Droplet size distributions at varying turbulence intensities shown in terms of their probability density functions (color-mapped) for three initial droplet size distributions characteristic of polluted, relatively clean, and pristine environments, respectively. The insets show detailed probability density functions at lower ($Re_\lambda$ = 136) and higher ($Re_\lambda$ = 198) turbulence intensities. **a,** Polluted condition. Droplet size distributions remain unchanged across all turbulence intensities, with no measurable growth observed. **b,** Relatively clean condition. The droplet size distribution broadens with increasing turbulence intensity, with creation of larger droplets at the expense of smaller ones. The inset depicts a noticeable change in the droplet size distribution at higher turbulence intensity. **c,** Pristine condition. The droplet size distribution broadens substantially with increasing turbulence, showing formation of larger droplets at greater expense to smaller droplets. The inset confirms this. **d,** Normalised spacing of droplet pairs as a function of average Stokes numbers ($\bar{St}$). The radial distribution function $g(r)$ (which defines the probability of finding a droplet pair at spacing $r$) is shown in color. The vertical lines indicate three different Stokes numbers whose variations of $g(r)$ are shown in the inset. At $\bar{St}$ = 0.17 (polluted condition), the value of $g(r)$ is close to one, indicating no clustering. In contrast, for $\bar{St}$ = 1.2 (relatively clean condition) and for $\bar{St}$ = 2.5 (pristine condition), we observe high clustering at the dissipative scales, more pronounced in the latter.

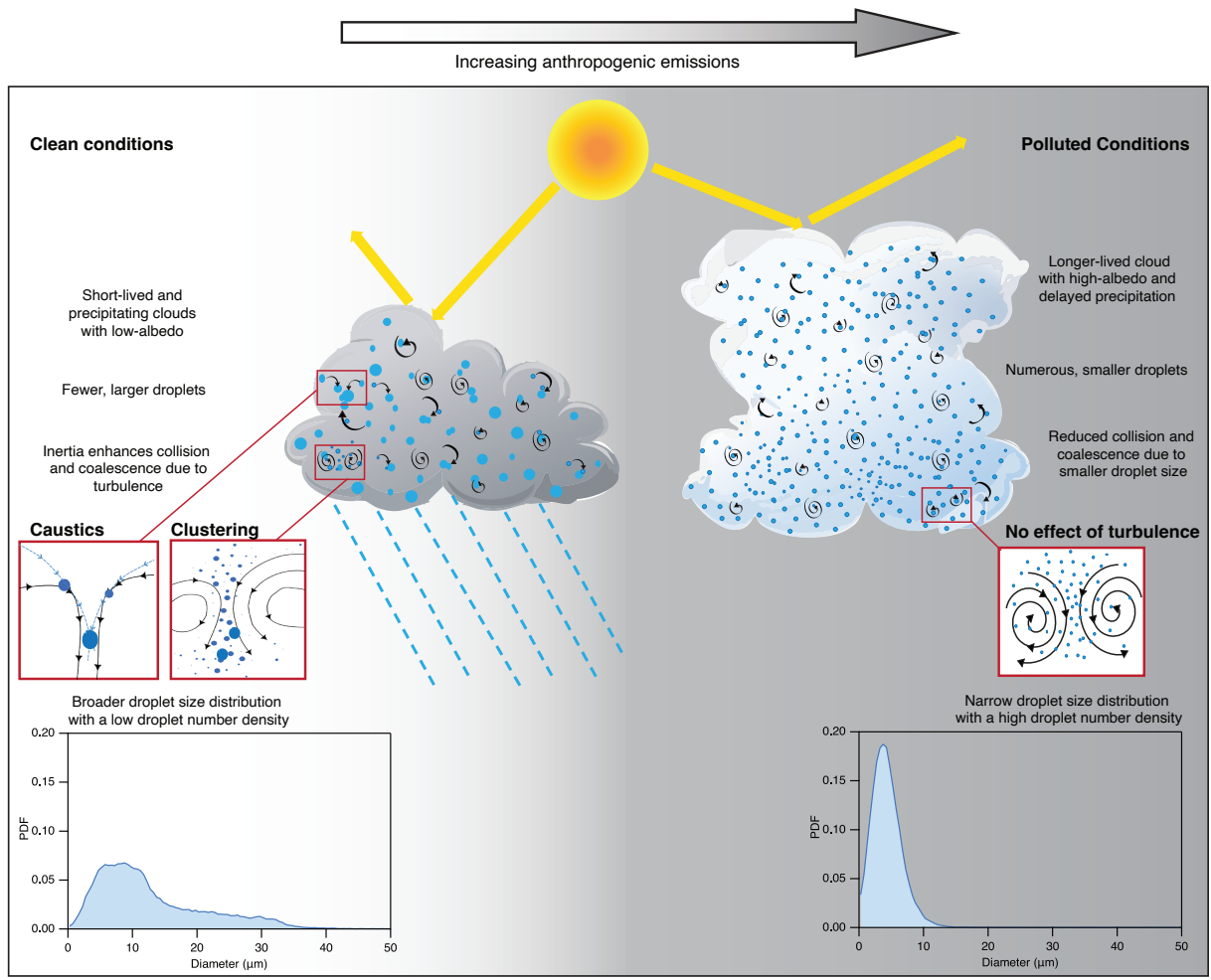

**Figure 2: In-cloud turbulence and characteristic cloud properties:** Characteristic clouds formed under cleaner conditions would have lower aerosol concentrations, resulting in broader size distributions composed of fewer, yet larger, droplets. The presence of in-cloud turbulent eddies and droplet inertia enhances clustering and caustics, causing droplets to collide and coalesce, which increases their size, and subsequently leads to bigger droplets that fall out as warm rain. Under the polluted conditions, the characteristic cloud would have smaller droplets with a narrow droplet size distribution and a high number concentration. Unlike the larger droplets, these smaller droplets follow the flow streamlines, thereby reducing the chance of collision and coalescence. Thus, the in-cloud turbulence has no significant influence on cloud droplet size growth for a condition with narrow droplet size distribution, resulting in an increased cloud lifetime and enhanced cloud albedo.

**Extended Data:**

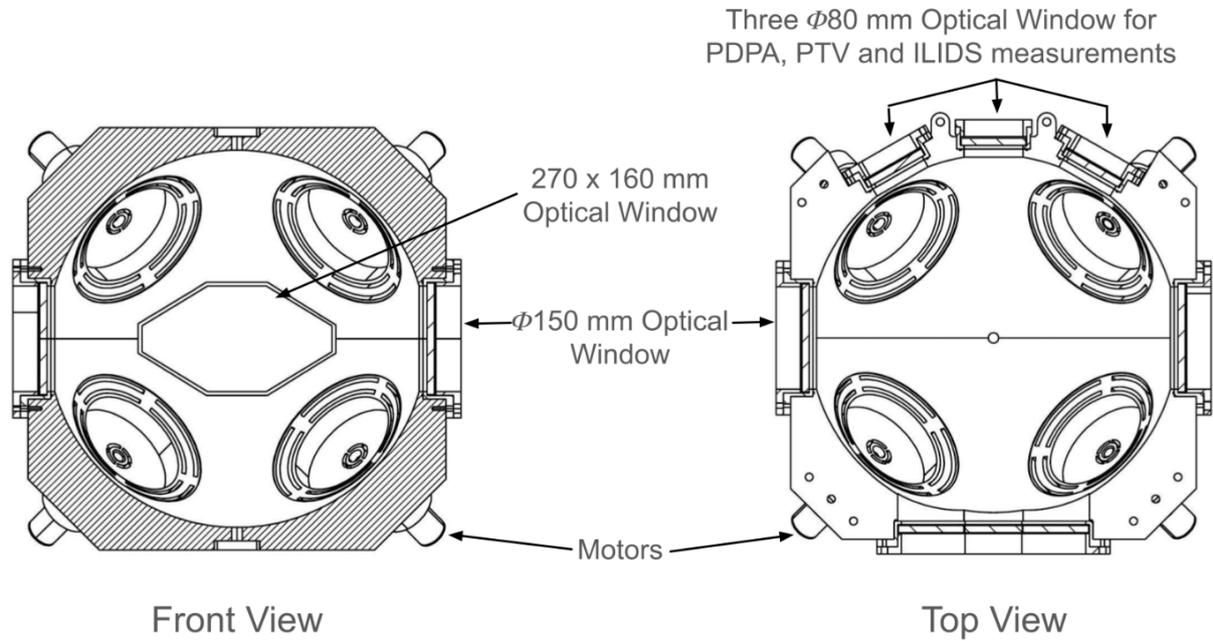

**Extended Data Fig. 1: Schematic shows the turbulence chamber dimensions (in mm).** The inside of the chamber has a spherical shape with an inner diameter of 500 mm, and the outside of the chamber is of a truncated cuboid shape. The chamber has six optical windows made of 10 mm-thick quartz for performing various optical diagnostic techniques. Eight 150 mm fans, coupled with 180 W brushless DC motors, were mounted symmetrically opposite to each other to drive the turbulence inside the chamber. Servo-motor speeds were regulated using digital controllers, ensuring operation in the same direction at closely matched speeds (within ±1% variation), thereby generating homogeneous, isotropic turbulence with zero-mean flow.

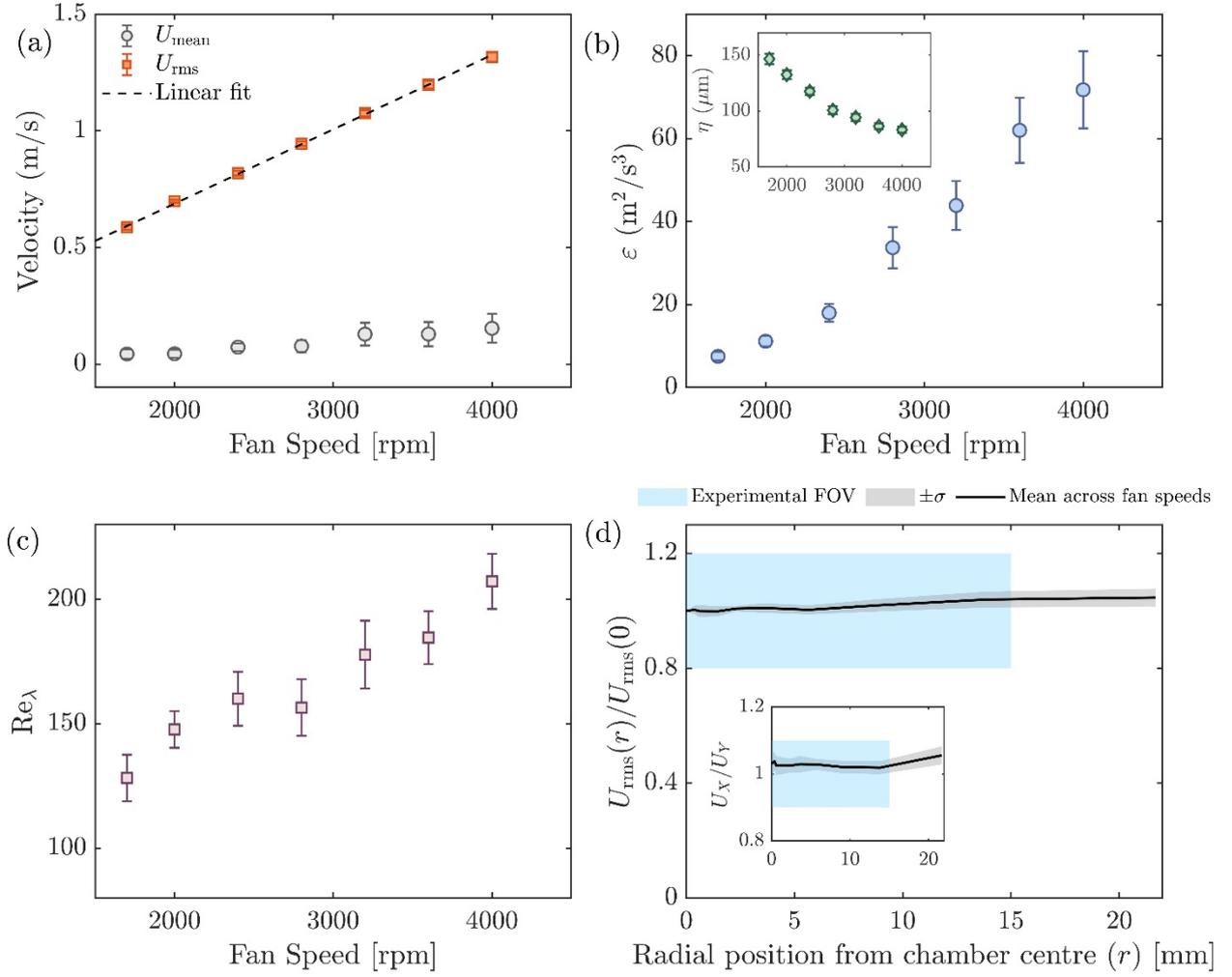

**Extended Data Fig. 2: Characterization of turbulence in the chamber.** **(a)** Root-mean-square velocity ($U_{rms}$) and mean velocity ($U_{mean}$) as a function of the fan speed. $U_{rms}$ increases approximately linearly with the fan speed, while $U_{mean}$ remains close to zero across all conditions. **(b)** Dissipation rate ($\epsilon$) as a function of fan speed (inset: corresponding Kolmogorov length scale $\eta$). **(c)** Variation of the Taylor-scale Reynolds number ($Re_\lambda$) with fan speeds. **(d)** Radial profiles of $U_{rms}$ and the velocity anisotropy ratio $U_{rms\_X}/U_{rms\_Y}$ demonstrate that the flow remains spatially homogeneous and near-isotropic (variations <10%) within the central 25 mm measurement region. The cyan shaded band marks the field of view used for the laser imaging of droplets.

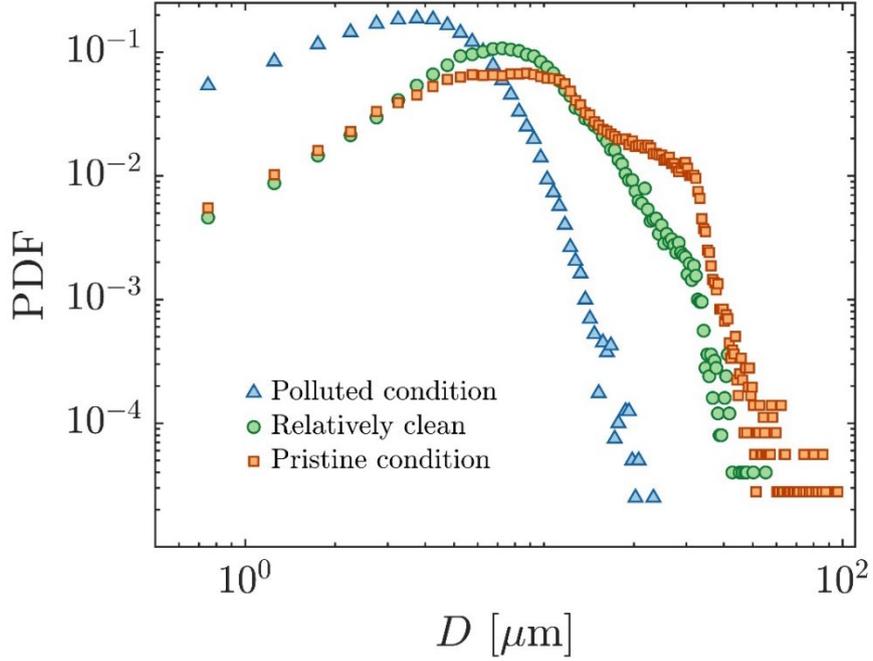

| Parameter | Polluted condition | Relatively clean | Pristine condition |
|---|---|---|---|
| Mean diameter (D10) μm | 4.3 | 9.4 | 12.8 |
| Relative dispersion (σ/D10) | 0.52 | 0.56 | 0.66 |
| Number concentration per cc | $O(1000)$ | $O(300)$ | $O(100)$ |

**Extended Data Fig. 3: Three characteristic droplet size distributions representing clouds originating from different environmental conditions.** Droplet size distribution (PDF) and parameters measured at the lower turbulence intensity ($Re_\lambda = 136$) are listed. Blue points correspond to polluted-cloud conditions, which exhibit a narrow distribution with high droplet number concentrations. Green points represent relatively clean-cloud conditions, showing a broader distribution at moderate concentrations. Red points denote pristine-cloud conditions, characterised by the broadest distributions and the lowest droplet concentrations.

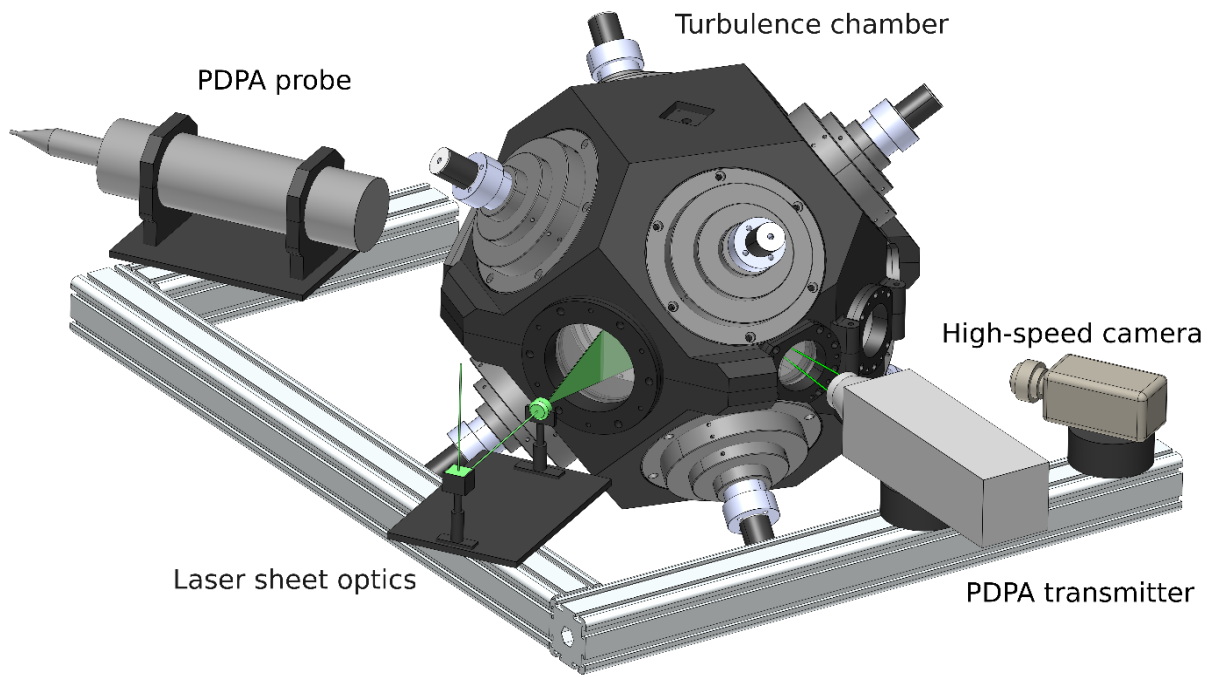

**Extended Data Fig. 4: Schematic of the experimental arrangement.** Diagram illustrating the overall configuration of the experimental facility, inducing the turbulence chamber, Phase Doppler Particle Analyzer (PDPA) system and the imaging system. The PDPA transmitter and receiver optics are aligned at a 33° off-axis scattering angle. The droplet field was illuminated using a laser sheet and imaged with a high-speed camera positioned orthogonally to the illumination plane.

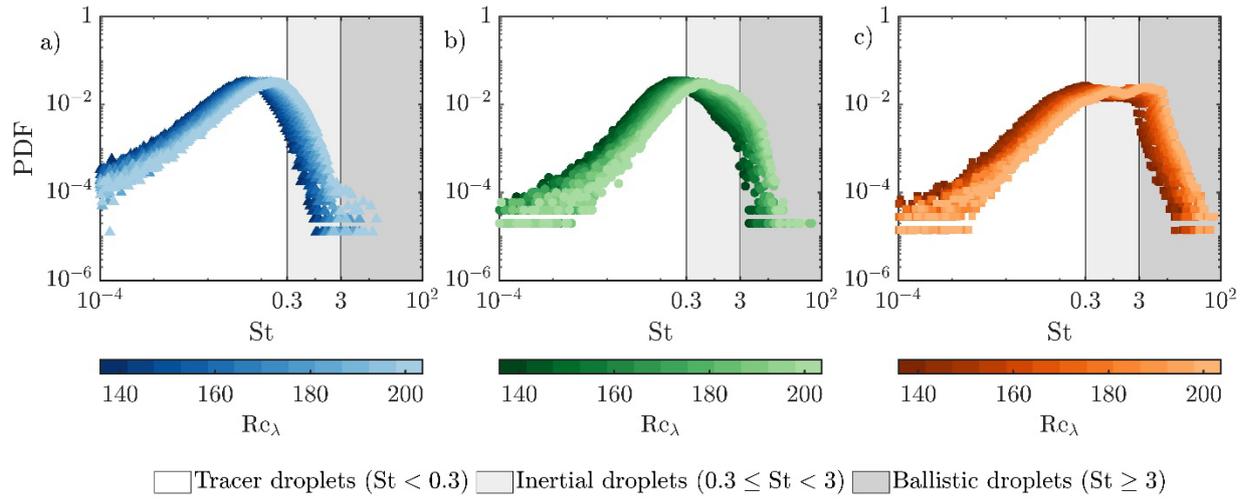

**Extended Data Fig. 5: Broad size distribution in pristine conditions produces droplets with wide range of St.** Plots (a), (b), and (c) show the distribution of St at different turbulence intensities for polluted, relatively clean, and pristine conditions respectively. An increase in turbulence intensity is marked by a decrease in the hue of the color. We group the droplets as tracers (St ≤ 0.3), inertial (0.3 < St ≤ 3), and ballistic droplets (St > 3). Case (3) has a wide range of St increasing interactions between droplets with disparate St. Also, increasing $Re_\lambda$ increases the St.

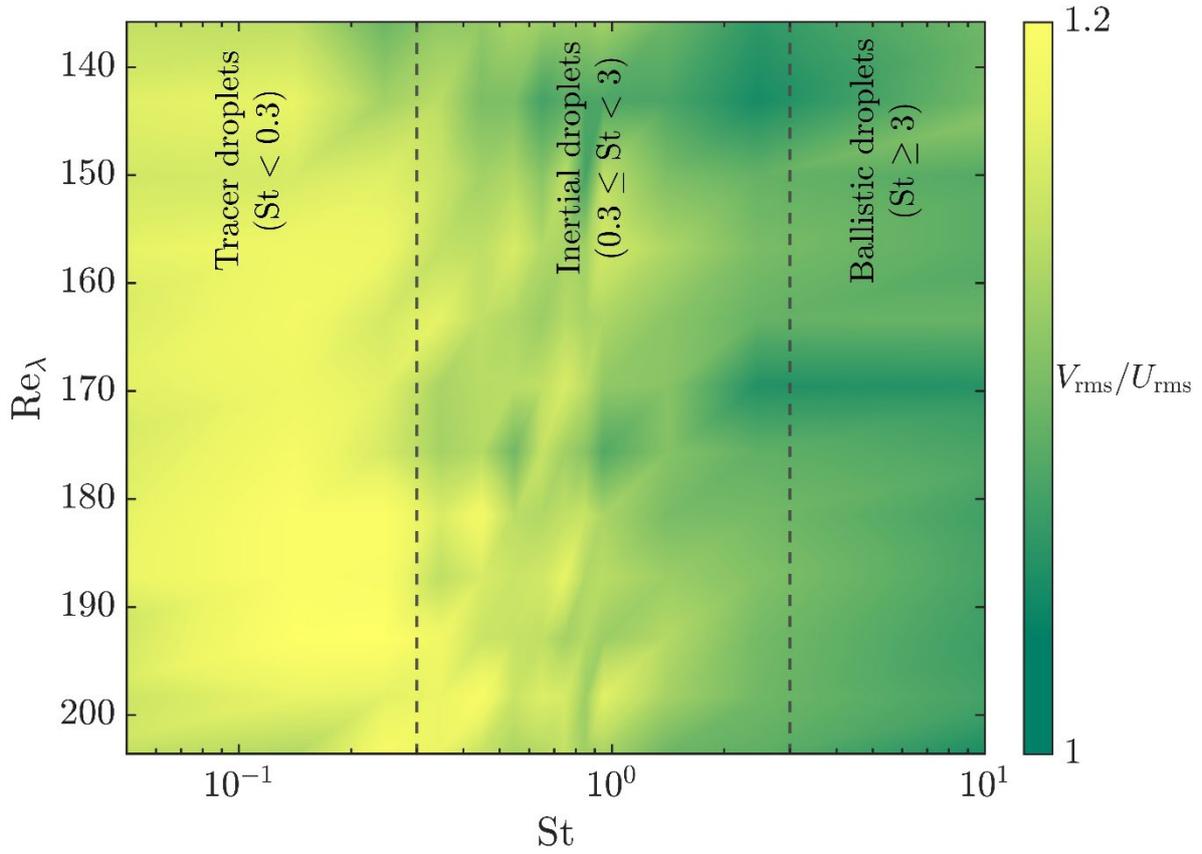

**Extended Data Fig. 6: Droplets with higher St move slower than rest, thereby enhancing caustic formation.** For droplets in pristine condition, we observe droplets with higher St move slower than rest, indicating a possibility of enhanced caustic formation, under the influence of turbulence.